\documentclass{bmcart}

\usepackage{amsthm,amsmath}
\RequirePackage{hyperref}
\usepackage[utf8]{inputenc} 

\def\includegraphics{}

\startlocaldefs
\endlocaldefs

\usepackage{graphicx}

\begin{document}

\begin{frontmatter}

\begin{fmbox}
\dochead{Research Note}


\title{AutoClassWeb: a simple web interface for Bayesian clustering}


\author[
  addressref={aff1},                   
  corref={aff1},                       
  email={pierre.poulain@u-paris.fr}   
]{\inits{P.}\fnm{Pierre} \snm{Poulain}}
\author[
  addressref={aff1},
]{\inits{J-M.}\fnm{Jean-Michel} \snm{Camadro}}


\address[id=aff1]{
  \orgname{Université de Paris, CNRS, Institut Jacques Monod},        
  \city{F-75013, Paris},                      
  \cny{France}                                
}



\end{fmbox}


\begin{abstractbox}

\begin{abstract} 
\parttitle{Objective} 
Data clustering is a common exploration step in the omics era, notably in genomics and proteomics where many genes or proteins can be quantified from one or more experiments. Bayesian clustering is a powerful algorithm that can classify several thousands of genes or proteins. AutoClass C, its original implementation, handles missing data, automatically determines the best number of clusters but is not user-friendly.

\parttitle{Results} 
We developed an online tool called AutoClassWeb, which provides an easy-to-use web interface for Bayesian clustering with AutoClass. Input data are entered as TSV files. Results are provided in formats that ease further analyses with spreadsheet programs or with programming languages, such as Python or R. AutoClassWeb is implemented in Python and is published under the 3-Clauses BSD license. The source code is available at \url{https://github.com/pierrepo/autoclassweb} along with a detailed documentation.
\end{abstract}




\end{abstractbox}
%

\end{frontmatter}



\section*{Introduction}

In biology, high-throughput technologies (notably in genomics and proteomics) enable identification and quantification of several thousands of genes or proteins in a single experiment. To analyze such a large amount of data, from one or more experiments, clustering algorithms are widely used unsupervised machine-learning methods to group genes or proteins with similar patterns. Bayesian clustering is such an algorithm and one of its implementation in the C programming language (AutoClass C) has been developed in 1996 at the Ames Research Center at NASA \cite{cheeseman1996,stutz1996}.
The idea behind Bayesian clustering and the AutoClass algorithm is to find 
a classification that fits the data with the highest probability. 
The AutoClass algorithm provides some additional and interesting features:
it handles missing data and determines automatically the best number of clusters.

AutoClass C has been used in a wide variety of applications from clustering cells of the prefrontal cortex in rats and mice \cite{elliott2018} to detecting body patterns in the common cuttlefish \cite{crook2002} (see also references \cite{achcar2009} and \cite{camadro2019} for a detailed list of applications). However, AutoClass C, originally developed by physicists, is not user-friendly: the program is solely accessible through the command line, only 32-bit binaries are available and results files are difficult to parse for subsequent analysis.

More than 10 years ago, Achcar \textit{et al.} published AutoClass@IJM \cite{achcar2009},
a web interface for AutoClass C. This web service drastically simplified the use of AutoClass C and widen its adoption, especially in biology \cite{elliott2018,wu2015,leger2015,franco2019,duval2020}. Unfortunately, this tool is not maintained anymore, and its source code is not publicly available.

To continue to offer this powerful Bayesian clustering method to the community, we developed AutoClassWeb, a new easy-to-use open-source web interface for AutoClass C.

\section*{Main text}

\subsection*{Implementation}

AutoClassWeb utilizes AutoClassWrapper \cite{camadro2019}, a Python wrapper for the AutoClass C program. This wrapper facilitates the preparation and quality control of data, runs the actual classification, and eventually, prepare results in file formats that allow further analysis.

AutoClassWeb is written in Python \cite{vanrossum1995} and uses the Flask library to build the web interface users interact with. For better reproducibility and sustainability, AutoClassWeb is
packaged in a Docker image stored in the BioContainers \cite{daveigaleprevost2017} registry.

The web service itself has been designed to be user-friendly. There is no user authentication and by default, results are kept 30 days before being deleted. A comprehensive \textit{help} page provides all the help and guidance the user might need.

Using Docker technology, AutoClassWeb can be quickly deployed on a local machine or on a public web server. To this end and to reduce the installation burden, we provide two companion GitHub
repositories with detailed instructions, for local (\url{https://github.com/pierrepo/autoclassweb-app}) and server installation (\url{https://github.com/pierrepo/autoclassweb-server}).

\subsection*{Data submission}

The input data must be formatted as tab-separated values (TSV) files. The first line is a header containing the names of the columns which must be unique.
The first column contains the names of the objects studied (\textit{i.e.} protein or gene identifiers).

Missing data is supported and should be coded with an empty value (\textit{i.e.} nothing)..

AutoClass supports three categories of data:

\begin{itemize}
	\item \emph{real location}: negative and positive values such as position, elevation, microarray log ratio...
	\item \emph{real scalar}: singly bounded real values, typically bounded below at zero (\emph{i.e.}: length, weight, age).
	\item \emph{discrete}: qualitative data. For instance, color, phenotype, name...
\end{itemize}

If the initial input dataset contains several data types (\textit{real scalar}, \textit{real location}, \textit{discrete}), it is recommended to split the initial dataset into several datasets of homogeneous type and submit them in the input form (Figure~\ref{fig:app-views} (A)).

For the data types \textit{real scalar} and \textit{real location}, the user can optionally specify an absolute and relative error, respectively.

\subsection*{Clustering}

\begin{figure}[!h]
	\centerline{\includegraphics[width=12cm]{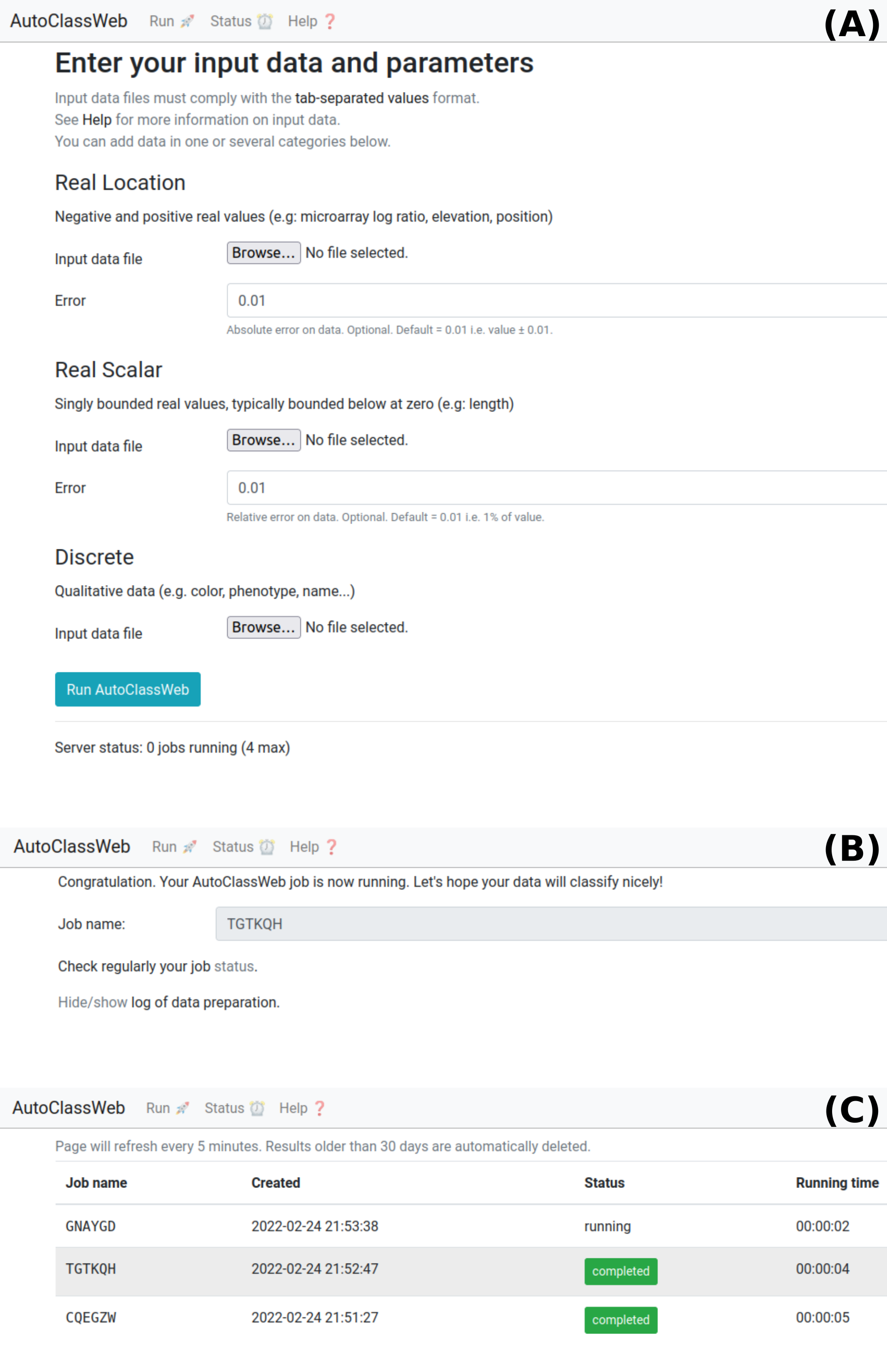}}
	\vspace*{-0.5cm}
	\caption{Views of AutoClassWeb. (A) Main page with the form to input data according to its type. (B) View after data input and quality check. (C) Status page.}\label{fig:app-views}
\end{figure}

Upon submission, input data is quality checked and formatted to be usable by AutoClass C.
The web interface provides a unique job name, a link to the status page and a quick summary of input data (toggled with the text \texttt{Hide/show logs}), as illustrated in Figure~\ref{fig:app-views} (B).

The status page lists running, failed and completed runs with their respective identifier 
(\textit{Job name}), creation date, status and running time (Figure~\ref{fig:app-views} (C)).

\subsection*{Results}

Once a job is completed, a green button allows the user to download results of the clustering. Results are bundled in a zip archive with the following files (where \texttt{xxx} stands for the unique identifier of the job):

\begin{itemize}
	\item \texttt{xxx\_autoclass\_out.cdt} and \\ \texttt{xxx\_autoclass\_out\_withproba.cdt} can be viewed with Java TreeView \cite{saldanha2004}, a versatile viewer initially developed for microarray data.
	The file \texttt{xxx\_autoclass\_out\_withproba.cdt} contains the probability for each object (gene or protein) to belong to each class.
	\item \texttt{xxx\_autoclass\_out\_stats.tsv} contains means and standard deviations of numeric columns (\textit{real scalar} and \textit{real location} data types) for each class.
	\item \texttt{xxx\_autoclass\_out\_dendrogram.png} is a dendrogram plot that visualizes the distance between all classes.
	\item \texttt{xxx\_autoclass\_out.tsv} contains all the data with the class assignment and membership probabilities for all classes. 
	This file is in the TSV format and can be easily parsed with spreadsheet programs such as Microsoft Excel or LibreOffice Calc, or programming languages such as R or Python.
\end{itemize}

\subsection*{Performance}

\begin{figure}[!h]
	\centerline{\includegraphics[width=12cm]{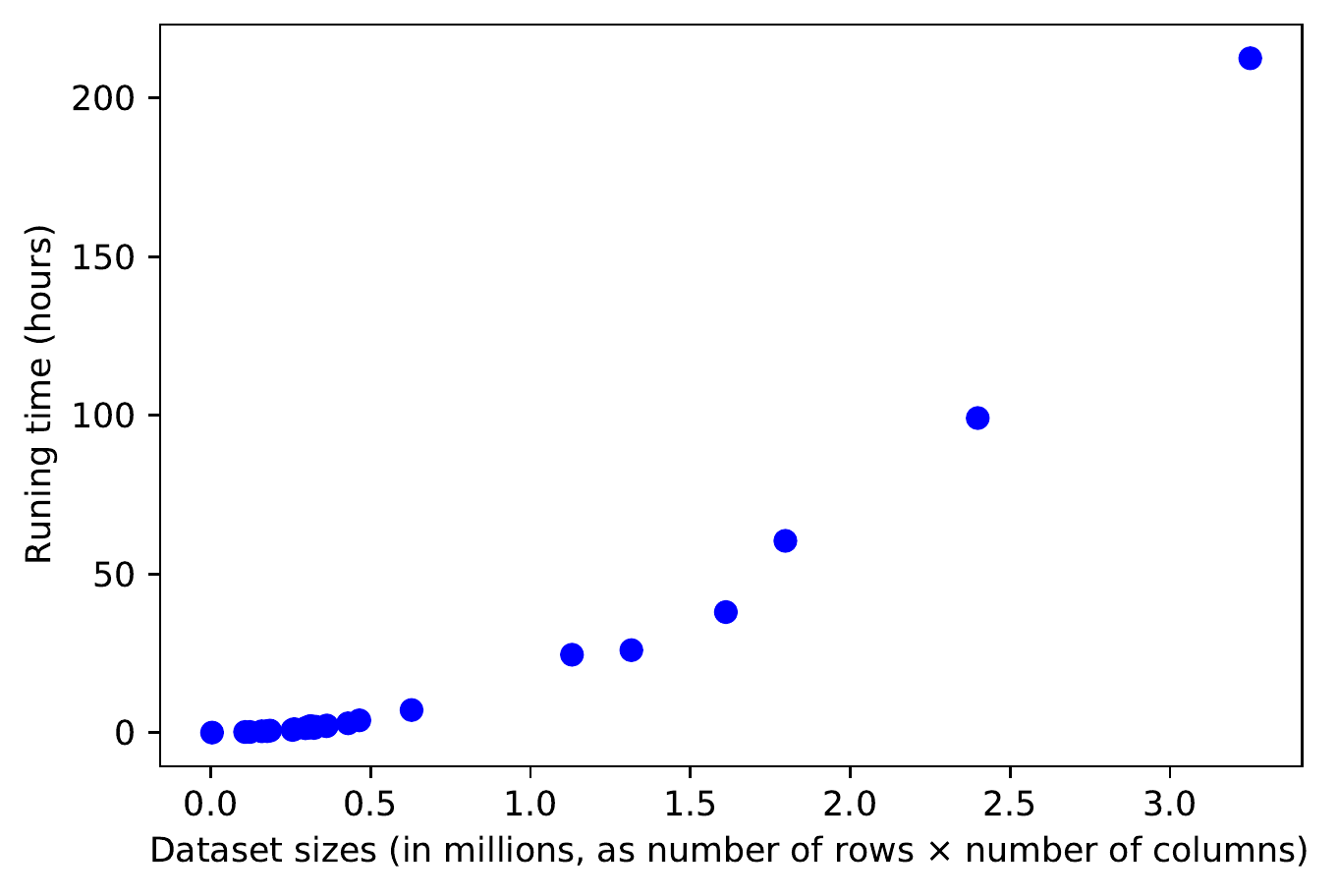}}
	\vspace*{-0.5cm}
	\caption{Running time (in hour) as a function of the input dataset sizes. Dataset size is expressed as the number of rows (genes or proteins) times the number of columns (features or studied properties).}\label{fig:running-time}
\end{figure}

The AutoClass C algorithm has been designed to run on a single CPU. The running time depends exponentially on the size of the input dataset. 
Figure \ref{fig:running-time} illustrates the running time as a function of the input dataset sizes. Dataset size is expressed as the number of rows (usually genes or proteins) times the number of columns (features or properties of interest).

\subsection*{Conclusion}

Data clustering is an essential step in most modern omics analyses. The AutoClass algorithm, while very powerful, is not widely used, mainly because its original AutoClass C implementation is difficult to use. AutoClassWeb provides an easy-to-use web interface for AutoClass C. The project is open-source, packaged in a Docker image available in BioContainers for better reproducibility and sustainability.

\section*{Limitations}

AutoClassWeb provides a convenient online service to cluster results from high-throughput experiments such as RNA-seq or mass spectrometry based proteomics. However, we would like to point out that the processing time required to cluster data with AutoClass is proportional to the number of genes or proteins to be clustered. A parallel version of AutoClass C that potentially reduces the processing time has been published \cite{pizzuti2003}. Unfortunately, the source code is not available, and the project has been discontinued.

Another limitation of AutoClassWeb requires users to split input data by type (\textit{real location}, \textit{real scalar} or \textit{discrete}) with a special attention to \textit{real location} and \textit{real scalar} which may sometimes be confused.


\bibliographystyle{vancouver} 
\bibliography{paper}      

\begin{thebibliography}{10}

\bibitem{cheeseman1996}
Cheeseman P, Stutz J.
\newblock Bayesian {{Classification}}({{AutoClass}}):{{Theory}} and
  {{Results}}.
\newblock In: Fayyad UM, Piatetsky-Shapiro G, Smyth P, Uthurusamy R, editors.
  Advances in {{Knowledge Discovery}} and {{Data Mining}}. {AAAI/MIT Press};.
  p. 153--180.

\bibitem{stutz1996}
Stutz J, Cheeseman P.
\newblock Autoclass - {{A Bayesian Approach}} to {{Classification}}.
\newblock In: Skilling J, Sibisi S, editors. Maximum {{Entropy}} and {{Bayesian
  Methods}}. No.~70 in The {{Fundamental Theories}} of {{Physics}}: {{Their
  Clarification}}, {{Development}} and {{Application}}. {Kluwer Academic
  Publishers};. Available from:
  \url{https://link.springer.com/book/10.1007%2F978-94-009-0107-0}.

\bibitem{elliott2018}
Elliott MC, Tanaka PM, Schwark RW, Andrade R.
\newblock Serotonin {{Differentially Regulates L5 Pyramidal Cell Classes}} of
  the {{Medial Prefrontal Cortex}} in {{Rats}} and
  {{Mice}};5(1):ENEURO.0305--17.2018.

\bibitem{crook2002}
Crook AC, Baddeley R, Osorio D.
\newblock Identifying the Structure in Cuttlefish Visual
  Signals;357(1427):1617--1624.

\bibitem{achcar2009}
Achcar F, Camadro JM, Mestivier D.
\newblock {{AutoClass}}@{{IJM}}: A Powerful Tool for {{Bayesian}}
  Classification of Heterogeneous Data in Biology;37:W63--W67.

\bibitem{camadro2019}
Camadro JM, Poulain P.
\newblock {{AutoClassWrapper}}: A {{Python}} Wrapper for {{AutoClass C}}
  Classification;4(39):1390.

\bibitem{wu2015}
Wu S, Clevenger JP, Sun L, Visa S, Kamiya Y, Jikumaru Y, et~al.
\newblock The Control of Tomato Fruit Elongation Orchestrated by Sun, Ovate and
  Fs8.1 in a Wild Relative of Tomato;238:95--104.

\bibitem{leger2015}
Léger T, Garcia C, Ounissi M, Lelandais G, Camadro JM.
\newblock The {{Metacaspase}} ({{Mca1p}}) Has a {{Dual Role}} in
  {{Farnesol-induced Apoptosis}} in {{{\emph{Candida}}}}{\emph{
  Albicans}};14(1):93--108.

\bibitem{franco2019}
Franco M, Vivo JM.
\newblock Cluster {{Analysis}} of {{Microarray Data}}.
\newblock In: Bolón-Canedo V, Alonso-Betanzos A, editors. Microarray
  {{Bioinformatics}}. vol. 1986 of Methods in {{Molecular Biology}}. {Springer
  New York};. p. 153--183.

\bibitem{duval2020}
Duval C, Macabiou C, Garcia C, Lesuisse E, Camadro J, Auchère F.
\newblock The Adaptive Response to Iron Involves Changes in Energetic
  Strategies in the Pathogen {{{\emph{Candida}}}}{\emph{ Albicans}};9(2).

\bibitem{vanrossum1995}
van Rossum G. Python Tutorial [Technical Report];.

\bibitem{daveigaleprevost2017}
da~Veiga~Leprevost F, Grüning BA, Alves~Aflitos S, Röst HL, Uszkoreit J,
  Barsnes H, et~al.
\newblock {{BioContainers}}: An Open-Source and Community-Driven Framework for
  Software Standardization;33(16):2580--2582.

\bibitem{saldanha2004}
Saldanha AJ.
\newblock Java {{Treeview--extensible}} Visualization of Microarray
  Data;20(17):3246--3248.

\bibitem{pizzuti2003}
Pizzuti C, Talia D.
\newblock P-Autoclass: Scalable Parallel Clustering for Mining Large Data
  Sets;15(3):629--641.

\end{thebibliography}



\begin{backmatter}

\section*{Acknowledgements}
Authors thank Denis Mestivier for fruitful discussions on AutoClass@IJM.

\section*{Availability of data and materials}
The source code of AutoClassWeb is open-source, released under the BSD-3-Clause license, and available on the GitHub development platform
\url{https://github.com/pierrepo/autoclassweb}.

AutoClassWeb source code in its current version 2.2.1 is archived in Software Heritage with the following reference: 
\href{https://archive.softwareheritage.org/swh:1:dir:173e846a5137a4b498ca7da9eca980790631bc1a;origin=https://github.com/pierrepo/autoclassweb;visit=swh:1:snp:47a3f95b96258f62d406c1cfc403505f19a5c95a;anchor=swh:1:rel:c8d2a2c08a600017d10b4b7ec492faa293c2514d}{swh:1:dir:173e846a5137a4b498ca7da9eca980790631bc1a}

\section*{Abbreviations}

\textbf{CPU}: Central Processing Unit

\textbf{TSV}: tab-separated values

\section*{Competing interests}
The authors declare that they have no competing interests.

\section*{Authors' contributions}
PP and JMC contributed to the design of the web interface. PP developed the tool and wrote the manuscript. JMC extensively tested the ergonomy. PP and JMC reviewed the manuscript. All authors have read and approved the final manuscript.

\end{backmatter}



\section*{Figures}

\end{document}